\documentclass[proof]{WileyASNA-v1}

\newcommand{\Msol}{\mathrm{M_\odot}}
\newcommand{\be}{\begin{equation}}
\newcommand{\ee}{\end{equation}}
\usepackage{orcidlink}

\articletype{Article Type}%

\received{26 April 2016}
\revised{6 June 2016}
\accepted{6 June 2016}

\begin{document}

\title{Gravitational waves or X-ray counterpart? No need to choose}% \protect\thanks{This is an example for title footnote.}}

\author[1]{Raphaël Mignon-Risse*}

\author[2,3]{Peggy Varniere}

\author[2]{Fabien Casse}

\authormark{Mignon-Risse \textsc{et al}}

\address[1]{Université Paris Cité, CNRS, CNES, Astroparticule et Cosmologie, F-75013 Paris, France}

\address[2]{Université Paris Cité, CNRS, Astroparticule et Cosmologie, F-75013 Paris, France}

\address[3]{Université Paris-Saclay, Université Paris Cité, CEA, CNRS, AIM, 91191, Gif-sur-Yvette, France}

\corres{*Raphaël Mignon-Risse\,\orcidlink{0000-0002-3072-1496}, \email{raphael.mignon-risse@apc.in2p3.fr}}

%\presentaddress{This is sample for present address text this is sample for present address text}

\abstract{Binary black holes emit gravitational waves as they inspiral towards coalescence. Searches for electromagnetic counterparts to these gravitational waves rely on looking for common sources producing both signals. In this paper, we take a different approach: we investigate the impact of {radiation zone effects, including retardation effects and gravitational wave propagation} onto the circumbinary disk around stellar-mass, spinning black holes, using general relativistic hydrodynamical simulations. Then we used a general relativistic ray-tracing code to extract its X-ray spectrum and lightcurve. This allowed us to show that {radiation zone effects} leave an imprint onto the disk, leading to quasi-periodic patterns in the X-ray lightcurve. The amplitude of the modulation is weak (${<}1\%$) but increases with time and is strongly dependent on the inclination angle.}


\jnlcitation{\cname{%
\author{R. Mignon-Risse}, 
\author{P. Varniere}, and 
\author{F. Casse}} (\cyear{2023}), 
\ctitle{Gravitational waves or X-ray counterpart? No need to choose}, \cjournal{Astronomical Notes}}

%%\fundingInfo{Funding info text.}

\maketitle

\section{Introduction}

Binary black holes (BBHs) emit gravitational waves (GWs) when inspiralling towards coalescence.
These are detectable with the current interferometers LIGO/Virgo/Kagra and the future LISA mission during part of the inspiral phase.
While a joint GW-electromagnetic (EM) observation has been obtained from the merger of two neutron stars \citep{abbott_gravitational_2017} together with a short gamma-ray burst, such a co-detection is still lacking for BBHs.
An EM counterpart would give us insights on the environment around the new populations of astrophysical objects that are stellar-mass BBHs.
In particular, the behaviour of the plasma in such time-varying spacetimes is uncertain.
Furthermore, comparing the time of arrival of photons and GWs would pose new constraints on the speed of gravity, as a test of General Relativity.
Moreover, since gravity dominates other forces in the BBH vicinity, one can expect the accretion flow to share common properties between stellar-mass and supermassive BBHs.
Gaining insight on those accretion flows would be beneficial before the launching of LISA in order to prepare the synergies between LISA and future EM facilities (e.g. ATHENA).
Such synergies offer additional scientific products such as the measure of distances of cosmological scales.

The lack of EM counterpart detection so far already questions whether stellar-mass BBHs are surrounded by an environment prone to produce an EM counterpart.
Theoretically, under particular conditions, they could harbor a gas-rich accretion environment.
First of all, stellar-mass BBHs are expected to form from the collapse of binary massive stars, which is accompanied by a supernova explosion. 
There, a light (${<}1\Msol$) fallback disk following supernova explosion of stellar material in rotation could provide this gas-rich environment (e.g. \citealt{chan_impact_2020}), with even the possibility to be a luminous accretion/ejection system \citep{mineshige_black_1997}.
Second, the tidal disruption of an interstellar cloud or star \citep{rees_tidal_1988} may lead to disk formation around the stellar-mass BBH.
Hints of TDE disks have been reported in observations {(e.g. \citealt{lin_large_2017})} and the subsequent formation of accretion structures studied numerically (e.g. \citealt{bonnerot_disc_2016}).
Even though TDEs are facilitated by the large tidal radius of massive BHs, the large number of stellar-mass BBHs in the Galaxy \citep{lamberts_predicting_2018} could be an asset to expect TDEs from stellar-mass BBHs in young stellar clusters (\citealt{kremer_fast_2020}, see also \citealt{ryu_close_2022}). 
Up to now, stellar-mass BBHs harboring a post-TDE disk may not be distinguishable from a low-mass X-ray binary system, i.e. a system hosting only one compact object.
Finally, another possibility is the presence of BBHs in the disk of active galactic nuclei.
This scenario has gained interest in the past few years (see e.g. \citealt{ford_binary_2021}, \citealt{li_hot_2021}) because the time to merger there would be considerably shortened in comparison with quiescent galactic nuclei \citep{mckernan_ram-pressure_2019}.

In case of a gas-rich environment around the stellar-mass BBH, a circumbinary disk is expected to form (from angular momentum conservation).
{In this paper, we study how the circumbinary disk behaves in the radiation zone where GWs will propagate and retardation effects become important (see \citealt{mignon-risse_impact_2022}), that we will refer to as radiation zone effects for conciseness, looking for its spectro-timing properties as seen by a distant observer.}
An EM signal directly related to the {radiation zone effects} would point, unequivocally, to a BBH system rather than a single BH with the same mass and disk.

For the following, we use the geometric system of units, i.e. $\mathrm{G}=\mathrm{c}=1$.
In this system, $1 \mathrm{M_\odot}=1.477\mathrm{km}=4.926\times 10^{-6}$~s.

\section{Impact of gravitational waves propagating through a disk}
\label{sec:gw}

To study the influence of {radiation zone effects} on a circumbinary disk, we computed the temporal evolution of a circumbinary disk with the GRAMRVAC code \citep{casse_impact_2017} implementing the BBH approximate analytical spacetime \citep{ireland_inspiralling_2016} {valid for spinning, non-eccentric, BBHs}.

\subsection{Approximate spacetime around a spinning binary black hole}

In a nutshell, the spacetime construction describes the gravitational impact of a spinning, non-precessing BBH in the GW-dominated inspiral phase.
The spacetime in the vicinity (the so-called near zone, NZ) of the BBH is described with a {post-Newtonian} metric, valid in the slow-motion ($v/\mathrm{c} \ll 1$, where $v$ is the BBH velocity and $\mathrm{c}$ the speed of light), weak-field ($r_A/M \gg 1$, where $r_A$ is the distance from the $A-$th BH and $M$ the BBH total mass) approximation.
Further away from the BBH than a gravitational wavelength (${\sim}  \pi (r_{12}/M)^{3/2}$, with $r_{12}$ their orbital separation), the so-called {radiation (or wave, or far) zone} spacetime is obtained by direct integration of the relaxed Einstein’s equations \citep{johnson-mcdaniel_conformally_2009}.
The spacetime construction is valid up to the $2.5$~post-Newtonian order, i.e. including terms up to $(v/\mathrm{c})^{2.5}$.
The inspiral trajectory is obtained by solving the equation of motion similarly to \cite{combi_superposed_2021}.
Inspiral motion is driven by the emission of GWs and tidal heating (\citealt{alvi_energy_2001}, \citealt{ajith_ninja-2_2012}, \citealt{blanchet_gravitational_2014}).
We cover the inspiral phase from an orbital separation of $15$~M down to $8$~M, i.e. about $10$ orbital periods, i.e. until the Post-Newtonian approximation breaks down.

In order to compute the influence of {radiation zone effects} on a circumbinary disk, we solve the equations of general-relativistic hydrodynamics with the GRAMRVAC code.
We will select a region between $300$~M and $2100$~M, i.e. further way from the BBH than a gravitational wavelength{, a necessary distance to study radiation zone effects}.
{By accounted for the outer disk only, we either assume that waves produced in the innermost regions are damped enough when reaching the outer disk, or that the disk formation is on-going and that there is no inner disk yet.
Observationally, the outer disk contributes to the energy spectrum at a lower energy than the inner disk.}
In this study we only considered the equal-mass case, representative of most stellar-mass BBH populations drawn from LIGO/Virgo/Kagra runs (see e.g. \citealt{the_ligo_scientific_collaboration_population_2022}).
We varied the symmetric spin parameter $\chi_\mathrm{S}=(a_1+a_2)/2$ from $-0.9$ to $0.9$, where $a_A$ is the spin of the $A-$th BH,.
For more details on the numerical setup of the simulations presented here, see ~\ref{app}.

\subsection{Gravitational waves leave an imprint on the density}
\label{sec:density}

\begin{figure}[t]
	\centerline{\includegraphics[width=16pc,height=36pc]{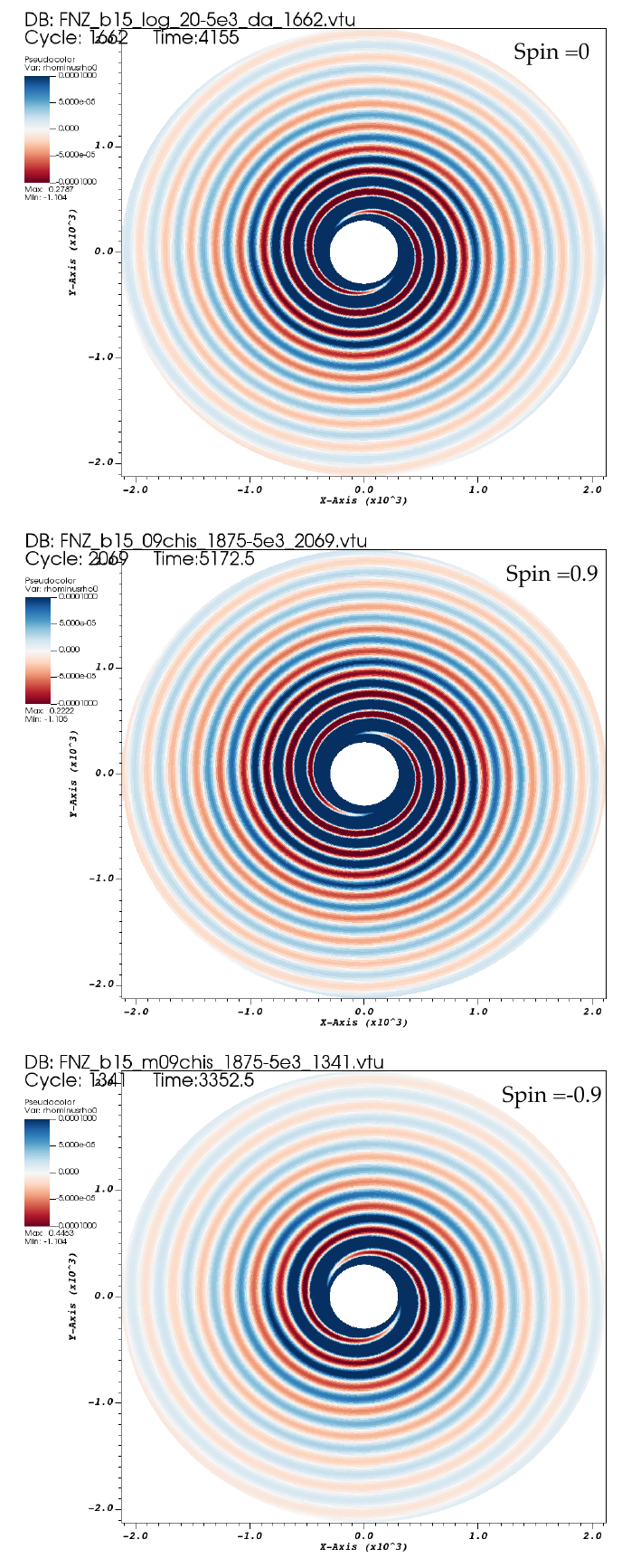}}
%	\centerline{\includegraphics[width=18pc,height=12pc]{Figures/FNZ_rhominusrho0.pdf}}
%	\centerline{\includegraphics[width=18pc,height=12pc]{Figures/FNZ_rhominusrho0_09chis_t5172.pdf}}
%	\centerline{\includegraphics[width=18pc,height=12pc]{Figures/FNZ_rhominusrho0_m09chis_t3352.pdf}}
	\caption{Spatial map of the deviation from the initial density distribution, for various spin parameters, when the orbital separation reaches $8$~M (the BBH is located within the inner boundary of the grid). 
	Top panel: $\chi_\mathrm{S}=0$. Middle panel: $\chi_\mathrm{S}=0.9$. Bottom panel: $\chi_\mathrm{S}=-0.9$. 
	A spiral structure is visible in the three maps.}
	\label{fig:FNZ_rhominusrho0}
\end{figure}

First of all, we look at how the plasma reacts to the metric not being Kerr but being that of {the radiation zone}.
Since we start from an axisymmetric disk that would be close to equilibrium in an equivalent Kerr metric, we are interested in the deviation of the density distribution with respect to the initial density distribution.
To do so, we show in Figure~\ref{fig:FNZ_rhominusrho0} the quantity $\rho(t_\mathrm{8M})-\rho(t=0)$, where $t_\mathrm{8M}$ stands for the time at which the orbital separation reaches $8$~M.
It shows that the GWs leave an imprint on the density distribution because it exhibits a non-axisymmetric, spiral-like structure, regardless of the BBH spin. 
We found the wavelength of the spiral to be increasingly shorter for increasing spins because it took more binary cycles to reach this orbital separation.
Let us note that this difference is visible since the first iterations of the simulation because the retardation effect is taken into account in the {radiation zone} and therefore accounts for the past history of the BBH.

\subsection{Temporal evolution of the impact}

\begin{figure}[t]
	\centerline{\includegraphics[width=20pc,height=10pc]{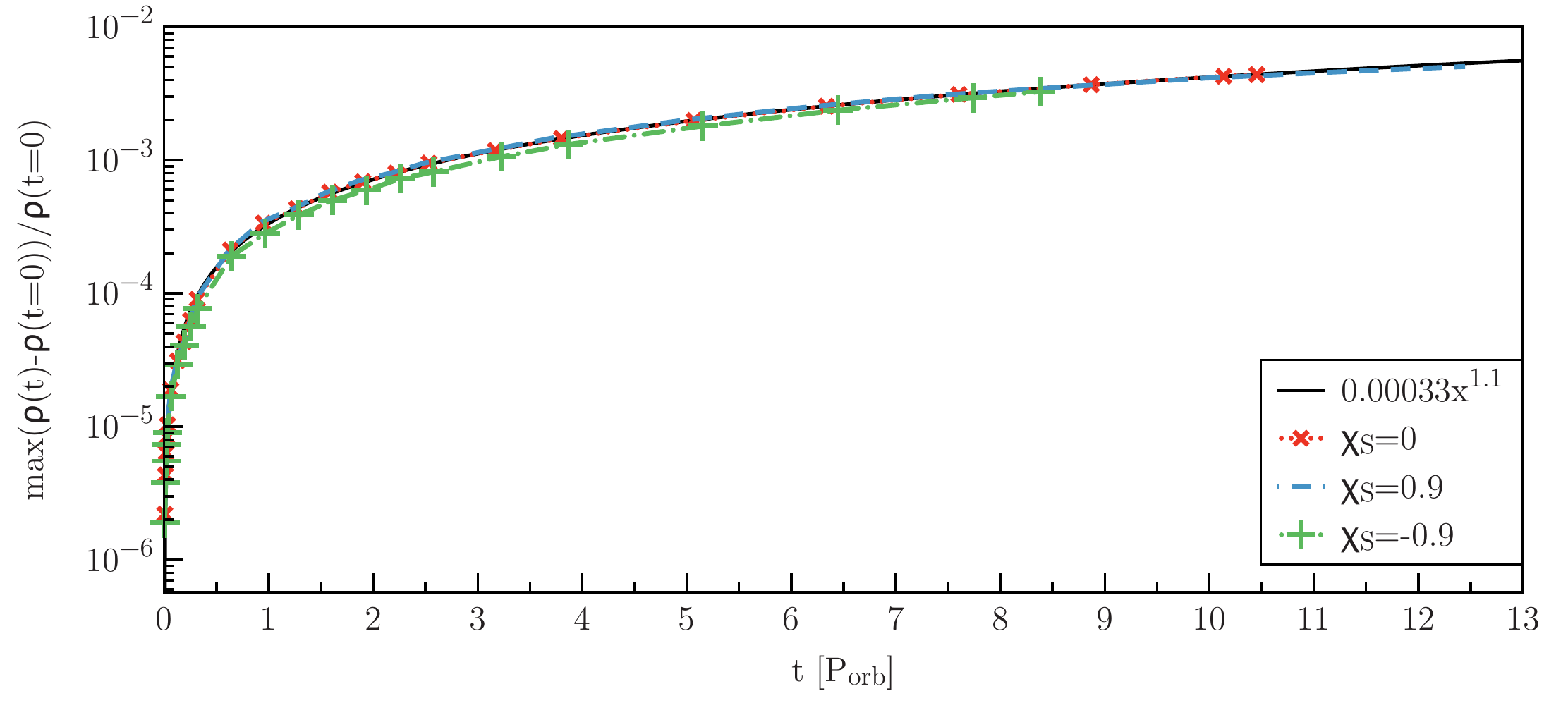}}%Figures/FNZ_t_rhominusrho0_2}}
	\caption{Temporal evolution (in units of the orbital period) of $\max(\rho(t)-\rho(t=0))/\rho(t=0)$ to illustrate the increase in amplitude of the spiral structure. 
	Cases $\chi_\mathrm{S}=0$ (Red dotted line and crosses), $\chi_\mathrm{S}=0.9$ (blue dashed line) and $\chi_\mathrm{S}=-0.9$ (green dashed-dotted line and "+" symbols) are shown, together with
	a power-law fit to the curve $\chi_\mathrm{S}=0$ after $3$ orbital periods (black curve).
	The simulations cover a BBH trajectory from $r_{12}=15$~M to $r_{12}=8$~M.
	The long-term increase of $\max(\rho(t)-\rho(t=0))/\rho(t=0)$ is similar for the various spin parameters, and, as shown by the fit, is consistent with a nearly linear increase.}
	\label{fig:FNZ_t_rhominusrho0}
\end{figure}

Since we have shown an impact of {radiation zone effects} on the density distribution of the circumbinary disk, we aim to quantify it and to search for its temporal evolution.
Figure~\ref{fig:FNZ_t_rhominusrho0} displays $\max(\rho(t)-\rho(t=0))/\rho(t=0)$ as a function of time, to see how the amplitude of the spiral grows with time.
This Figure shows that the amplitude of the spiral increases very similarly for these three values of $\chi_\mathrm{S}$, and that this increase is nearly linear, as indicated by the power law fit (black curve).
We show that, for $\chi_\mathrm{S}=0.9$, the effect acting on the density distribution can be integrated over a longer timescale than for $\chi_\mathrm{S}<0.9$, while having a similar growth rate (the slope).
At the time we ended the simulation, it reached a value of $0.5\%$.
%In order to give an alternative view on the initial small changes in $\rho$, we also show the same plot on a semi-logarithmic scale, zooming-in on the time interval covering the first orbital period.
%This shows how similar is the increase in amplitude of the spiral among the spin values during the very first stages of the simulation.
%This plot shows that the increase becomes compatible with the linear fit only after roughly $0.3$ orbital periods, suggesting that the early evolution is dominated by other processes than the long-term ones (e.g. relaxation with respect to the initial conditions).

To sum up, we found that a metric consistent with outgoing GWs leaves an imprint on the system's circumbinary disk which would be, otherwise, close to equilibrium.
In the following, we test how this imprint translates into EM observables.

\section{Associated X-ray observables}
\label{sec:x}

We have shown an impact of {radiation zone effects} on the density distribution of the circumbinary disk, which increases with time.
In order to test whether this could affects the EM observables, the simulations outputs need to be post-processed with a ray-tracing code.
Hence, we implemented the same BBH spacetime as before into the \textsc{GYOTO} code \citep{vincent_gyoto_2011} and ray-traced the photons from the simulated source to the observer.

To begin with, the ray-tracing step requires to re-dimensionalize the system.
Indeed, up to here, the results are normalized by the total mass of the BBH and thus valid for stellar-mass as well as for supermassive BBHs.
In this section, we choose to present synthetic lightcurves of the thermal emission associated to the circumbinary disk around an equal-mass, spinning, $40\Msol$ BBH (e.g. the GW191215 223052 event, \citealt{the_ligo_scientific_collaboration_gwtc-3_2021}).
The same values of the spin parameters $\chi_\mathrm{S}$ are used here as in the simulations presented above.
Let us note though that the temporal behaviour is - as shown in the simulations - independent on the mass of the object (hence it will not influence the behaviour of the frequency-integrated lightcurve, only its normalization), and will shift the energy at which the spectrum peaks.

We took the temperature at the innermost stable circular orbit to be $1$~keV, in agreement with spectral fits from low-mass X-ray binaries.
The overall flux is dominated by the flux between $0.01$ and $0.5$~keV, hence we only show soft X-ray in the following.
The flux is of the order of $10^{-10}$ to $10^{-6} \mathrm{erg\, cm^{-2} \, s^{-1}}$ for a source at $10$~kpc.

\subsection{The X-ray counterpart: a modulated lightcurve of increasing amplitude}

\begin{figure}[t]
	\centerline{\includegraphics[width=20pc,height=12pc]{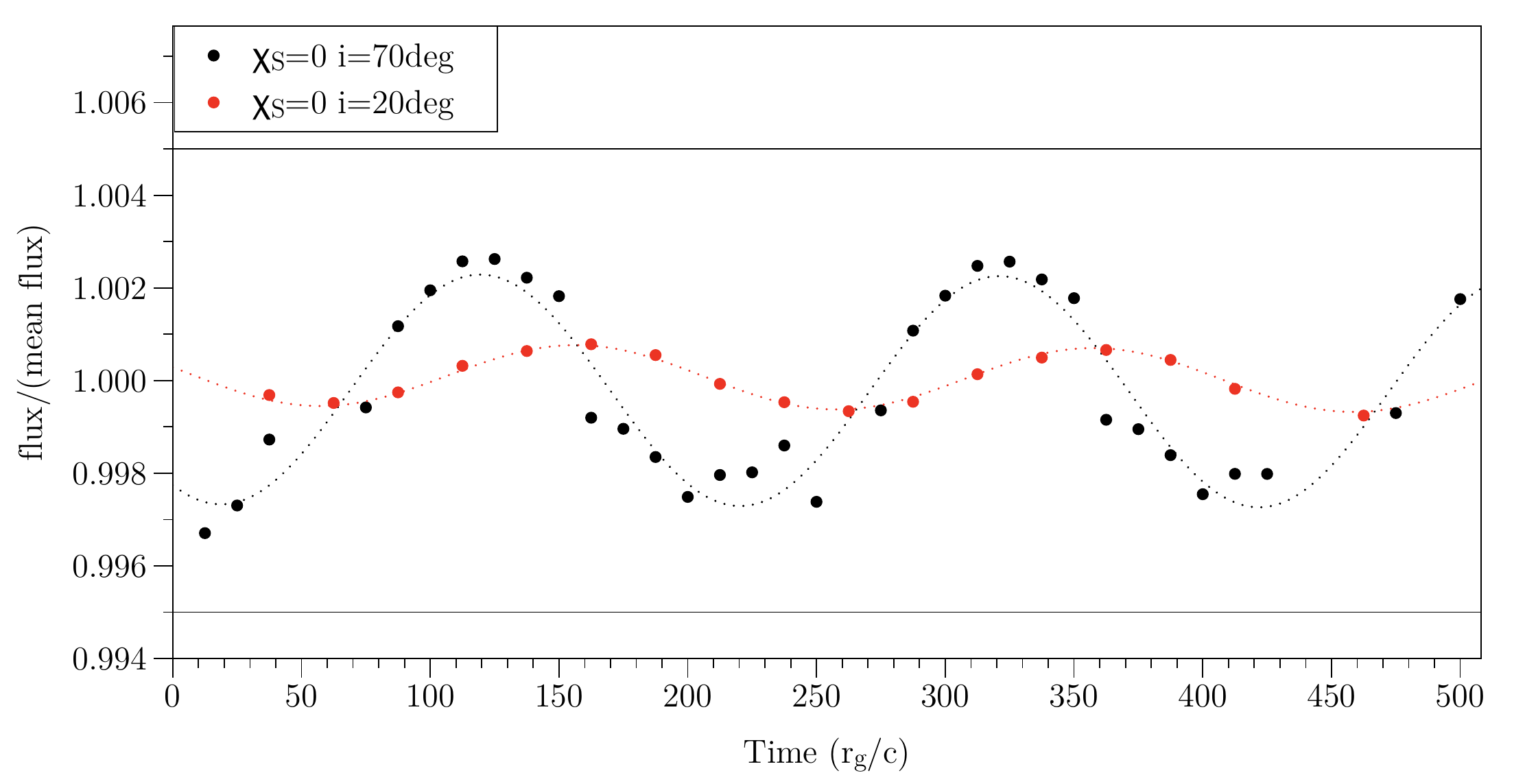}}%Figures/LC_FNZ_inclis.eps}}
	\caption{Lightcurve associated with the thermal emission from a circumbinary disk around an equal-mass, spinless, inspiralling BBH, for inclination angles of $20\deg$ and $70\deg$.
	The lightcurve is computed over roughly one orbital period.
	The lightcurve exhibits a weak (${<}1$\%) but clear modulation (indicated by the dotted curves) at the semi-orbital period.}
	\label{fig:LC_FNZ_inclis}
\end{figure}

We observed a spiral structure in the density distribution, whose amplitude increases with time, so we turned to the lightcurve to see whether it yields any timing feature.
Figure~\ref{fig:LC_FNZ_inclis} shows the lightcurve (here normalized by its mean value) associated with the thermal emission from the circumbinary disk in the spinless simulation for various inclinations, over roughly one orbital period{, e.g. ${\approx}391 \mathrm{r_g/c}=7.7\times 10^{-2} (r_{12}/15\mathrm{r_g})^{3/2} (M/40\mathrm{M_\odot})$~s}.
This figure shows that the lightcurve clearly exhibits a modulation at the semi-orbital period of the BBH.
The amplitude of the modulation is less than $1\%$.
As expected from the spiral structure in density the amplitude of the modulation is larger at higher inclination angle \citep{varniere_impact_2016}.

\begin{figure}[t]
	\centerline{\includegraphics[width=22pc,height=12pc]{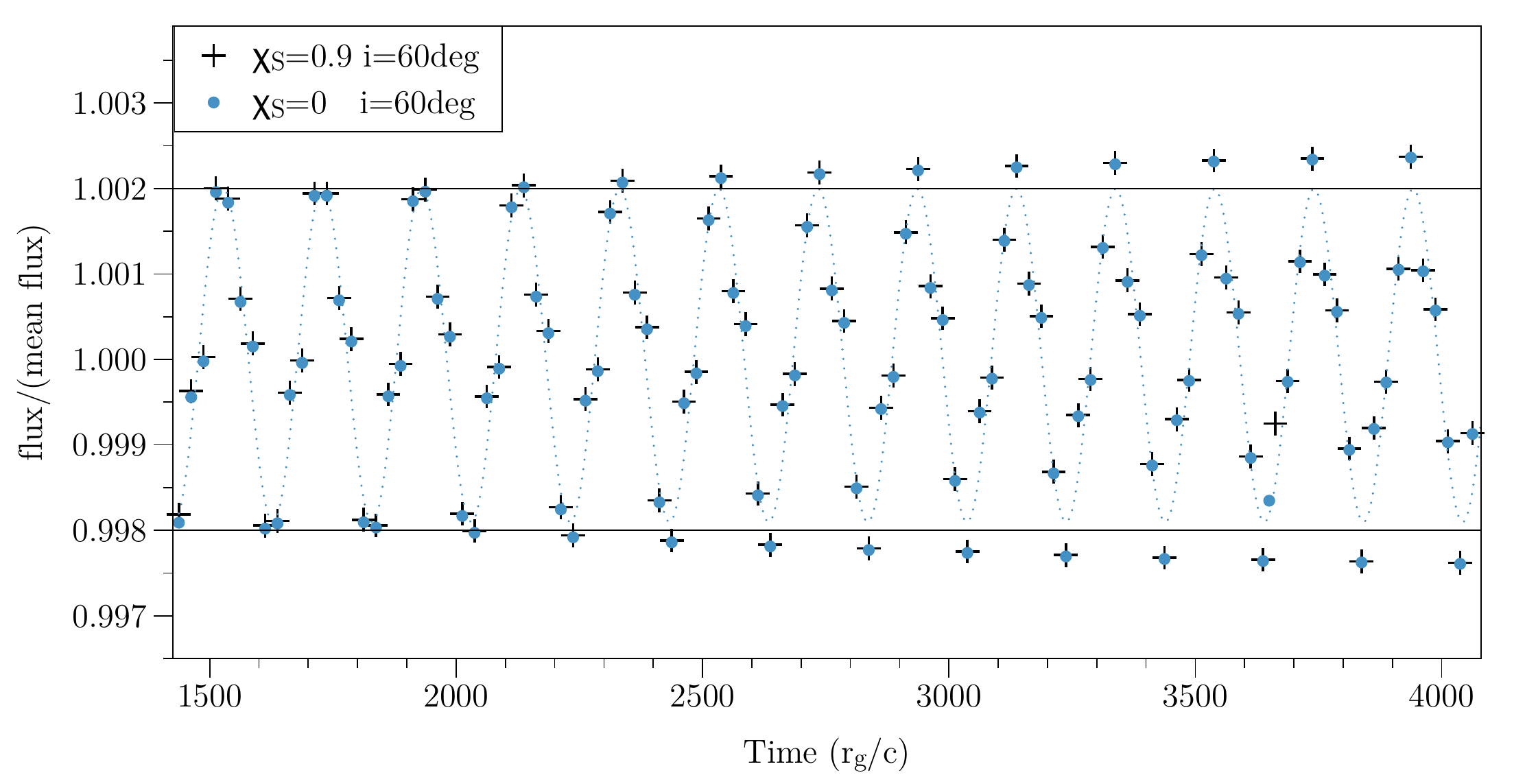}}%Figures/LC_long_term_spins.eps}}
	\caption{Lightcurve associated with the thermal emission from a circumbinary disk around an equal-mass, inspiralling BBH (runs $\chi_\mathrm{S}=0$ and  $\chi_\mathrm{S}=0.9$) over several orbits.
	The amplitude of the modulation of the lightcurve increases with time.
	As a visual guide, the dotted curve is a sinusoidal fit of the lightcurve on the first orbital period, extended to a larger number of periods.}
	\label{fig:LC_increase}
\end{figure}

In order to track the long-term behavior of this modulation, and motivated by the density distribution being increasingly impacted by {radiation zone effects} (Fig.~\ref{fig:FNZ_t_rhominusrho0}~), we computed the lightcurve over several orbits for $\chi_\mathrm{S}=0$ and $\chi_\mathrm{S}=0.9$.
The result is shown in Fig.~\ref{fig:LC_increase}.
The long-term lightcurve is characterized by an increase of the amplitude of the modulation.
It appears to be very similar for the two spin parameters considered here.
Let us also note that there is a slight decrease in the period of the modulation (consistent with the BBH trajectory computed) but it remains marginal and therefore hardly visible in this Figure.

An axisymmetric disk around a single BH would produce a flat lightcurve.
Hence, we suggest that this modulation has the potential to distinguish GW-emitting BBHs disks from axisymmetric disks around single BHs.

\subsection{Impact of the spin on the energy spectrum}

\begin{figure}[t]
	\centerline{\includegraphics[width=21pc,height=11pc]{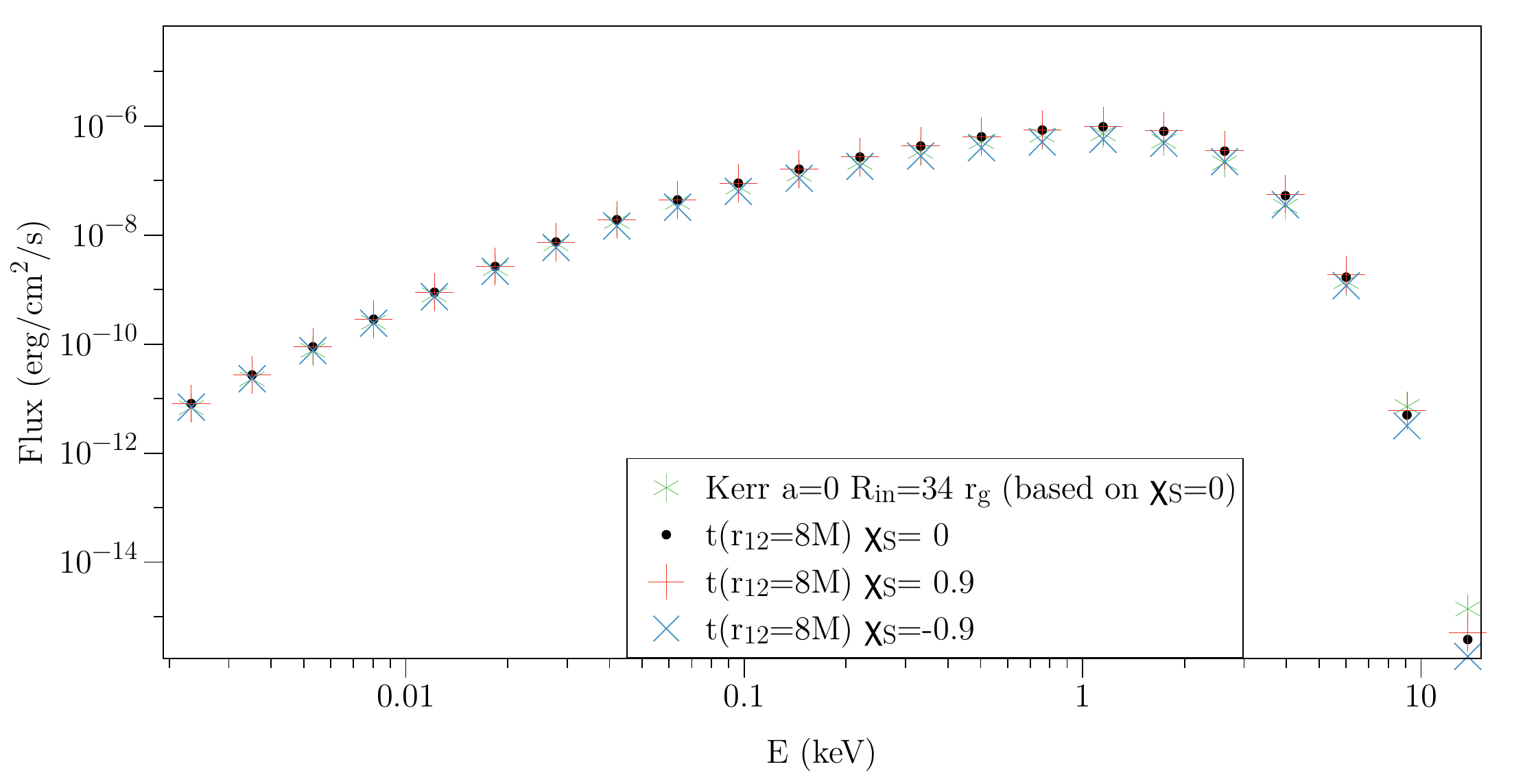}}%Figures/FNZ_spec_BBH_spins.eps}}
	\caption{Energy spectrum of a circumbinary disk around an equal-mass, $40\Msol$ BBH, for three values of the spin parameter $\chi_\mathrm{S}$. 
	The Kerr case with an inner disk radius such as to produce a comparable energy spectrum is shown.
	 The source inclination is $70\deg$.
	 First, the energy spectrum peaks in the soft X-ray band and is nearly indistinguishable between the different spin parameters.
	 Second, it does not allow to distinguish a BBH from a Kerr black hole.}
	\label{fig:spec}
\end{figure}

An impact of the BH spin is expected in the spectrum of low-mass X-ray binaries \citep{varniere_novas_2018}, hence we investigated a similar impact on the spectrum of the circumbinary disk (Figure~\ref{fig:spec}).
First, as mentioned above, the energy spectrum peaks in the soft X-ray band.
Moreover, we found that the energy spectrum is nearly invariant with the spin parameter $\chi_\mathrm{S}$, as expected from the lightcurves.
Consequently, it cannot be used to extract the spin parameter $\chi_\mathrm{S}$ of the source.
Finally, the difference with a Kerr BH is not detectable, so there is no spectral feature allowing to distinguish the BBH from a single BH.
From this study, focused on the outer parts of the circumbinary disk, we conclude that only a timing study provides potentially detectable differences with respect to a single BH disk.

\section{Conclusion}
\label{sec:ccl}

In this study, we have investigated the influence of {the combination of outgoing GWs and retarded potentials associated with} the inspiral of a BBH on the density distribution of the circumbinary disk.
To do so, we implemented a BBH analytical metric in the GR-(M)HD code (\textsc{GRAMRVAC}) and used it to follow the temporal evolution of the circumbinary disk density located in the far (radiation) zone.
With the aim to produce observables such as the lightcurve and the energy spectrum, in order to assess the observability of the impact on the circumbinary disk, we have post-processed the outputs of the simulation using a GR ray-tracing code (\textsc{GYOTO}).

We found that {radiation zone effects on} a circumbinary disk produce non-axisymmetric density structures with a spiral-like shape.
This effect is weak on the density distribution, with a deviation of up to $0.5\%$ with respect to the initial density (but it can be integrated for longer times, see below).
After computing the lightcurve, we found it to exhibit a modulation at the semi-orbital period because of the non-axisymmetry.
The amplitude of this modulation is less than $1\%$.
While low, this amplitude is in par with some of the weaker high-frequency quasi-periodic oscillations \citep{remillard_evidence_2002}, and therefore could be detectable.
The optimal case to detect such a modulation is that of a prograde spinning BBH seen at high inclinations.

The amplitude of the modulation reported here is a lower limit because the process under study actually begins at larger orbital separations and continues up to the merger, with a linear increase. 
However, following the temporal evolution of the circumbinary disk over the whole inspiral phase of the BBH is computationally prohibitive.

{Although we focused on stellar-mass BHs for being typical XMM sources, our results can be scaled to SMBBHs \citep{mignon-risse_impact_2022}.
An increase in BBH mass would reduce the emission frequency and increase the period of the signal, hence the variability coming from a stellar-mass BBH is widely different from a SMBBH.
}

Overall, we showed that {radiation zone effects} in a gas-rich medium, e.g. that of a fallback disk or a post-TDE disk, can allow us to detect BBHs.
Considering the shortness of the inspiral phase for stellar-mass BBHs - the present study only accounts for a portion of the trajectory of ${\sim}1(M/40 \Msol)$~second in the inspiral phase -, this X-ray counterpart would not be possible to detect on-the-fly.
However, it might be present in the EM archives associated with LIGO/Virgo sources.

\section*{Acknowledgments}

RMR thanks the anonymous referee for her/his comments.
RMR thanks Fabrice Dodu for optimizing the computation of the metric, Miguel Zillh\~ao and Hiroyuki Nakano for their insights on the equation of motion.
RMR acknowledges support from CNES through Postdoctoral fellowship.
%This work was supported by the Action Incitative "Ondes gravitationnelles et objets compacts" and by the Conseil Scientifique de l'Observatoire de Paris.
The numerical simulations we have presented in this paper were produced on the supercomputer Dante and on the HPC resources from GENCI-CINES (Grant A0100412463).
Part of this study was supported by the LabEx UnivEarthS, ANR-10-LABX-0023 and ANR-18-IDEX-0001.

%\subsection*{Author contributions}

%This is an author contribution text. This is an author contribution text. This is an author contribution text.  

%\subsection*{Financial disclosure}

%None reported.

%\subsection*{Conflict of interest}

%The authors declare no potential conflict of interests.

%\section*{Supporting information}

%The following supporting information is available as part of the online article:

%\noindent
%{Figure S1.}
%{500{\uns}hPa geopotential anomalies for GC2C calculated against the ERA Interim reanalysis. The period is 1989--2008.}

%\noindent
%{Figure S2.}
%{The SST anomalies for GC2C calculated against the observations (OIsst).}

\appendix

\section{Numerical setup}
\label{app}

In this appendix, we give numerical details on the simulations presented in the main text.

The initial conditions consist of an axisymmetric disk of density profile:
\be
\rho(r) = 0.5 \rho_0 \left( 1 - \tanh{ \left( \frac{r-2500}{500} \right) } \right) r^{-3/4} + \rho_\mathrm{min},
\ee
with $\rho_0=10$. We used a density threshold $\rho_\mathrm{min}=2 \times 10^{-4}$ in the initial density profile to avoid too small densities at large distances.
Thermal pressure is given by the equation of state $p=p_0 \rho^\gamma$, with the adiabatic index $\gamma$ set to $5/3$, and $p_0=1.8\times 10^{-4}$.
In order to track the influence of the BBH on its surrounding gas, we set the disk to be near equilibrium around an equivalent (same total mass) Kerr black hole, using the azimuthal velocity formula given in \cite{casse_impact_2017}.
The disk aspect ratio resulting from these choices of density, pressure and velocity varies between $0.1$ and $0.25$.

The grid inner and outer radii are $18.75$~M and $5000$~M (despite the disk outer edge being around $3000$~M), respectively, and the grid goes from $\phi=0$ to $\phi=2\pi$.
The resolution is set to $n_r \times n_\phi = 476 \times 140$, with a logarithmic spacing in $r$.
As a result, the finest resolution is $0.2$~M in the cavity (close to the BBH) and the resolution coarsens to be $20$~M in the far zone, i.e. roughly $20$ cells per gravitational wavelength.
For the boundary conditions, we set a continue boundary (i.e. null-gradient) at the radial inner and outer boundaries on $\rho$ and $v^i$, with an additional "outflow" condition to prevent material to enter the grid.

%\nocite{*}% Show all bib entries - both cited and uncited; comment this line to view only cited bib entries;
\bibliography{Zotero}%

\end{document}